\documentstyle[12pt, psfig]{article}
\newcommand{\nsect}{\setcounter{equation}{0}
\def\theequation{\thesection.\arabic{equation}}\section}

\catcode`\@=11
\def\marginnote#1{}
\def\ifmath#1{\relax\ifmmode #1\else $#1$\fi}

\def\bold#1{\setbox0=\hbox{$#1$}%
     \kern-.025em\copy0\kern-\wd0
     \kern.05em\copy0\kern-\wd0
     \kern-.025em\raise.0433em\box0 }

\def\GENITEM#1;#2{\par\vskip6pt \hangafter=0 \hangindent=#1
   \Textindent{$ #2$ }\ignorespaces}

\newcount\hour
\newcount\minute
\newtoks\amorpm
\hour=\time\divide\hour by60
\minute=\time{\multiply\hour by60 \global\advance\minute by-
\hour}
\edef\standardtime{{\ifnum\hour<12 \global\amorpm={am}%
    \else\global\amorpm={pm}\advance\hour by-12 \fi
    \ifnum\hour=0 \hour=12 \fi
    \number\hour:\ifnum\minute<100\fi\number\minute\the\amorpm}}
\edef\militarytime{\number\hour:\ifnum\minute<100\fi\number\minute}
\def\draftlabel#1{{\@bsphack\if@filesw {\let\thepage\relax
  \xdef\@gtempa{\write\@auxout{\string
    \newlabel{#1}{{\@currentlabel}{\thepage}}}}}\@gtempa
    \if@nobreak \ifvmode\nobreak\fi\fi\fi\@esphack}
     \gdef\@eqnlabel{#1}}
\def\@eqnlabel{}
\def\@vacuum{}
\def\draftmarginnote#1{\marginpar{\raggedright\scriptsize\tt#1}}
\def\draft{\oddsidemargin -.5truein
        \def\@oddfoot{\sl preliminary draft \hfil
        \rm\thepage\hfil\sl\today\quad\militarytime}
        \let\@evenfoot\@oddfoot \overfullrule 3pt
        \let\label=\draftlabel
        \let\marginnote=\draftmarginnote

\def\@eqnnum{(\theequation)\rlap{\kern\marginparsep\tt\@eqnlabel}%
\global\let\@eqnlabel\@vacuum}  }
\def\preprint{\twocolumn\sloppy\flushbottom\parindent 1em
        \leftmargini 2em\leftmarginv .5em\leftmarginvi .5em
        \oddsidemargin -.5in    \evensidemargin -.5in
        \columnsep 15mm \footheight 0pt
        \textwidth 250mmin      \topmargin  -.4in
        \headheight 12pt \topskip .4in
        \textheight 175mm
        \footskip 0pt

\def\@oddhead{\thepage\hfil\addtocounter{page}{1}\thepage}
        \let\@evenhead\@oddhead \def\@oddfoot{} \def\@evenfoot{}
}
\def\titlepage{\@restonecolfalse\if@twocolumn\@restonecoltrue\onecolumn
     \else \newpage \fi \thispagestyle{empty}\c@page\z@
        \def\thefootnote{\fnsymbol{footnote}} }
\def\endtitlepage{\if@restonecol\twocolumn \else  \fi
        \def\thefootnote{\arabic{footnote}}
        \setcounter{footnote}{0}}  
\catcode`@=12
\relax

\def\be{\begin{equation}}
\def\ee{\end{equation}}
\def\bea{\begin{eqnarray}}
\def\eea{\end{eqnarray}}
\def\simlt{\stackrel{<}{{}_\sim}}
\def\simgt{\stackrel{>}{{}_\sim}}

\def\NPB#1#2#3{{\it Nucl.~Phys.} {\bf{B#1}} (19#2) #3}
\def\PLB#1#2#3{{\it Phys.~Lett.} {\bf{B#1}} (19#2) #3}
\def\PRD#1#2#3{{\it Phys.~Rev.} {\bf{D#1}} (19#2) #3}

\def\PR#1#2#3{{\it Phys.~Rep.} {\bf#1} (19#2) #3}

\def\mst1{m_{\;\widetilde{t}_{1}}}

\def\mst2{m_{\;\widetilde{t}_{2}}}
\def\mst12{m_{\;\widetilde{t}_{1,2}}}

\def\MSbar{\overline{\rm MS}}

\def\msb1{m_{\;\widetilde{b}_{1}}}
\def\msb2{m_{\;\widetilde{b}_{2}}}
\def\msb12{m_{\;\widetilde{b}_{1,2}}}

\def\mtilde2{\widetilde{m}^{2}}

\relax

\begin{document}
\topmargin-2.5cm
\begin{titlepage}
\begin{flushright}
IEM-FT-200/00 \\ IFT-UAM/CSIC-00-10 \\ hep--ph/0002205 \\
\end{flushright}
\vskip 0.3in
\begin{center}{\Large\bf The Standard Model instability and
the scale of new physics~\footnote{Work supported in part by the CICYT
of Spain (contract AEN98-0816).} }
\vskip .5in {\large\bf J.A. Casas, V. Di Clemente and  M. Quir{\'o}s}
\\
\vskip.35in
{\large Instituto de Estructura de la Materia (CSIC)\\ Serrano 123,
28006-Madrid, Spain

\vspace{0.2cm}

Instituto de F\'{\i}sica Te\'orica, C-XVI\\ Universidad Aut\'onoma de
Madrid, 28049 Madrid, Spain }
\end{center}
\vskip1.5cm
\begin{center}
{\bf Abstract}
\end{center}
\begin{quote}
We apply a general formalism for the improved effective potential with
several mass scales to compute the scale $M$ of new physics which is
needed to stabilize the Standard Model potential in the presence of a
light Higgs.  We find, by imposing perturbativity of the new physics,
that  $M$ can be as large as one order of magnitude higher than the
instability scale of the  Standard Model.  This implies that, with the
present lower bounds on the Higgs mass, the new physics could easily
(but not necessarily) escape detection in  the present and future
accelerators.
\end{quote}

\vskip1.5cm

\begin{flushleft}
IEM-FT-200/00 \\ IFT-UAM/CSIC-00-10 \\ February 2000 \\
\end{flushleft}
\end{titlepage}
\setcounter{footnote}{0} \setcounter{page}{0}
\newpage

\nsect{Introduction}

The Standard Model (SM) effective potential is unstable beyond a
scale, $\Lambda_{SM}$, which solely depends upon the value of the (yet
undiscovered) Higgs mass $M_H$ \cite{sher1} ~\footnote{The original
studies    considered $\Lambda_{SM}$ as a function of $M_H$ and $M_t$
(the top-quark mass) \cite{SM}. In this paper we will fix $M_t$ to its
experimental mean value and disregard the effect due to the
experimental error $\Delta M_t$.}. This instability is a drawback of
the Standard Model, which is unable to describe physics at scales
beyond $\Lambda_{SM}$ and, then, requires the presence of new physics
to stabilize the SM vacuum. For large values of the Higgs mass the
scale $\Lambda_{SM}$ is larger than the Planck scale and thus has no
impact on the physics at present and future accelerators. However, for
the case of  a light Higgs (i.e. $M_H\simlt 140$ GeV) $\Lambda_{SM}$
can be closer to values that could directly, or indirectly, be
detected at future accelerators, thus having an impact on the physics
at the corresponding scales. In particular, and to guarantee the
absolute stability of the electroweak vacuum, new physics must be
introduced. The direct, or indirect, detection of the new physics
depends on the mass, $M$, of the new  involved fields, as deduced from
the stability requirement. This is the issue that will be considered
in this paper.

Since the presence of extra bosonic (fermionic) degrees of  freedom
tend to  stabilize (destabilize) the SM Higgs
potential~\cite{sher2, datta, neutrino2, neutrino1},   
it is clear that the simplest and
most economic SM extension that could  circumvent the instability
problem is the addition of just scalar fields coupled to the SM Higgs
field~\cite{sher2, datta}. This will be the model considered in the
present  paper. A similar  analysis of this kind of model was
performed by Hung and Sher in Ref.~\cite{sher2}, where the effective
potential was considered in the one-loop approximation. However
improving the effective potential by the renormalization group
equations (RGE) yields very important corrections and must be taken
into account in any realistic calculation. Obviously, other SM
extensions, such as the MSSM, can be much more realistic,  but, for
the previous reasons,  the simple model under consideration  basically
represents the most efficent way to cure the SM instability, thus
giving reliable upper bounds on the scale of new physics, i.e. on the
mass of the extra particles.

Improving the effective potential in a SM extension generally implies
considering a multi-scale problem. In our case, we have to consider
two different mass scales: the SM scale, say $\mu_t$, that can be
identified with the top-quark mass~\footnote{For the sake of
simplicity we will consider $\mu_t$ as the only SM scale, and will not
make a distinction with the other (nearby) SM scales, as the masses of
the Higgs and the gauge bosons.}, and the scale of new physics, say
$\mu_\Phi$, that should be identified with the mass of the new scalar
fields ($\Phi$). A general formalism to evaluate the effective
potential in  this kind of multi-scale scenario was presented by the
authors in  Ref.~\cite{cdq1}, where the general procedure for
decoupling fields~\cite{desacoplo} was  incorporated. We will use this
formalism throughout the paper.

The outline of the paper is as follows: In section \ref{multiscale} we
will briefly summarize our general proposal of multi-scale effective
potential. In section \ref{toy} we will apply the ideas and results of
section \ref{multiscale} to the model we are considering to remove the
SM instability. In section \ref{numerical} we will present our
numerical results and in section \ref{conclusions} our conclusions.
 
\nsect{Multi-scale effective theory}
\label{multiscale}

We will consider the effective potential of a theory with $N$
different scales in a mass-independent renormalization
scheme, as e.g. the $\overline{\rm MS}$ scheme, where the decoupling
is not automatically incorporated in the theory. The usual procedure
is to decouple every field at a given scale, $\mu_i$, that is normally
associated to its mass, by means of e.g. a step function. However, the
scale invariance of the complete effective potential indicates that
the results should not depend on the details of the chosen decoupling
scale $\mu_i$, thus implying independence of the (complete) effective
action with respect to the scale $\mu_i$ (similar to the usual scale
invariance).  This will give rise to a set of RGE with respect to the
scales $\mu_i$, as well as with respect to $\mu$, which is the
ordinary $\overline{\rm MS}$ renormalization scale. On top of that we
will use a simple step function for decoupling~\footnote{In addition,
at a given threshold the corresponding symmetry of the system may
change, in which case the corresponding matching conditions have to be
taken into account at the corresponding thresholds. A typical example
is the threshold of supersymmetric particles in the MSSM. Below it,
the theory is non-supersymmetric and beyond it supersymmetric. 
This kind of
complications will not appear, however, in the model at consideration
in the present paper.}, although it could be easily smoothed.  These
ideas were presented in Ref.~\cite{cdq1} and similar ones can be found in
Refs.~\cite{jones} and \cite{ford}.

In this decoupling approach, when computing the one-loop
$\beta$-functions corresponding to $\lambda_a$ [where $\lambda_a$
denotes all dimensionless and dimensional couplings of the theory,
including the  bosonic and fermionic fields],
the contribution from decoupled fields should not be counted. This
translates into the decomposition:
\begin{equation}
\mu\,\frac{d\, \lambda_a}{d\,\mu}\equiv\,_\mu\beta_{\lambda_a}
=\sum_{i=1}^N\,_{\mu_i}\widetilde\beta_{\lambda_a}\ \theta(\mu-\mu_i)
\label{betamu}
\end{equation}
which provides the definition of the factors
$_{\mu_i}\widetilde\beta_{\lambda_a}$.

Invariance of the complete effective potential with respect to the
$\MSbar$ scale $\mu$ simply reads
\begin{equation}
\label{invmu}
\mu\,\frac{d\,V_{\rm eff}}{d\,\mu}=0
\end{equation}
On the other hand, the one-loop effective potential can be written as:
\begin{equation}
\label{potencial}
V_{\rm eff}=V_0+V_1
\end{equation}
where $V_0$ is the tree-level potential, and
\begin{equation}
\label{effpot}
V_1=\frac{1}{64\pi^2} \sum_{i=1}^{N}(-)^{2s_i}
M_i^4\left[\log\frac{M_i^2}{\mu_i^2}+
\theta(\mu-\mu_i)\log\frac{\mu_i^2}{\mu^2}-C_i\right]
\end{equation}
where $s_i$ is the spin of the $i-th$ field, $M_i$ are the
(tree-level) mass eigenvalues and $C_i$ depends on the renormalization
scheme. In the $\MSbar$-scheme it is equal to 3/2 (5/6) for scalar
bosons and fermions (gauge bosons).  Notice that for $\mu>\mu_i$ (for
all $i$) the potential (\ref{effpot}) coincides with the usual
$\MSbar$ effective potential when there are no decoupled particles.
For $\mu<\mu_i$ the term $M_i^4\log(M_i^2/\mu_i^2)$ can be taken (at
the one-loop level) as frozen at the scale  $\mu=\mu_i$, so it does
not run with respect to $\mu$. On the other hand, the
invariance of the effective potential with respect to $\mu_i$,
\begin{equation}
\label{invmui}
\mu_i\,\frac{d\,V_{\rm eff}}{d\,\mu_i}=0
\end{equation}
leads to the running of the parameters with respect to the scale
$\mu_i$:
\begin{equation}
\mu_i\,\frac{d\, \lambda_a}{d\,\mu_i}\equiv\,_{\mu_i}\beta_{\lambda_a}
=\,_{\mu_i}\widetilde\beta_{\lambda_a}\ \theta(\mu_i-\mu)
\label{betamui}
\end{equation}
{}From Eqs.~(\ref{invmu}) and (\ref{betamui}) it follows that
\begin{equation}
\label{sumbeta}
_\mu\beta_{\lambda_a}+\sum_{i=1}^{N}\,_{\mu_i}\beta_{\lambda_a}=
\sum_{i=1}^{N}\,_{\mu_i}\widetilde\beta_{\lambda_a}=\beta_{\lambda_a}^{\MSbar}
\end{equation}
where $\beta_{\lambda_a}^{\MSbar}$ is the usual (complete)
$\beta$-function in the $\MSbar$-scheme.

The invariance of the effective potential with respect to $\mu$ and
$\mu_i$ allows to choose any values for these scales.  These can be
constant, as it is usually done~\cite{japoneses}, or 
field-dependent~\footnote{This is
similar to the ordinary $\MSbar$-scheme where the scale $\mu$ can be
fixed to a field dependent value.  In the case of the SM this value is
usually $\sim M_t$, in order to improve the validity of perturbation
theory.}.  A choice that is particularly convenient to greatly improve
the validity of perturbation theory is $\mu_i\simeq M_i$ and
$\mu\simlt\min{\{\mu_i\}}$ ~\cite{cdq1}.
%
%
This gets rid of all the dangerous logarithms in Eq.~(\ref{effpot}).
Notice here that since $M_i$ are in general functions of the fields,
so the preferred value of the  $\mu_i$ scales are.  In addition, the
evaluation of the effective potential requires to  evaluate all the
$\lambda_a$ parameters at the same values of $\mu_i$  (note that
$\lambda_a$ run with the $\mu_i$ scales). This implies a knowledge of
the $_{\mu_i}\widetilde\beta_{\lambda_a}$ functions defined in
Eq. (\ref{betamu}).

It is interesting to note that the previous decoupling approach can be
obtained starting with the bare lagrangian written in an appropriate
renormalization scheme. The latter is a generalization of the
so-called multi-scale renormalization scheme proposed in
Refs.~\cite{jones, ford}.  Given a set of bosonic $\phi_i$ and
fermionic $\psi_j$ fields, the bare lagrangian, in terms of
renormalized fields $(\phi_i,\psi_j)$ and renormalized couplings and
masses $\lambda_b$, can be written as:
\begin{equation}
\label{lagdec}
{\cal L}_{\rm Bare}=\sum_i {\cal L}_{{\rm kin},\,\phi_i}+ \sum_j {\cal
L}_{{\rm kin},\,\psi_j}+{\cal L}_{\rm Bare,\, int}
\end{equation}
where we have included in ${\cal L}_{\rm Bare,\, int}$ all interaction
and potential terms. More precisely,
\begin{eqnarray}
{\cal L}_{{\rm kin},\,\phi_i}&=&
\frac{1}{2}\widetilde\mu_i^{-\varepsilon/2}
f_{\phi_i}(\widetilde\mu_\ell)\,
Z_{\phi_i}\left(\partial_\mu\phi_i\right)^2 \nonumber\\ {\cal L}_{{\rm
kin},\,\psi_j}&=&i\, \widetilde\mu_j^{-\varepsilon/4}
f_{\psi_j}(\widetilde\mu_\ell)\, Z_{\psi_j} \overline\psi_j
\partial_\mu\gamma^\mu \psi_j \nonumber\\ {\cal L}_{\rm Bare,\,
int}&=& \sum_b\ Z_{\lambda_b}\  f_{\lambda_b}(\widetilde\mu_\ell)\
\lambda_b  O_b(Z_{\phi_i}^{1/2}\phi_i,Z_{\psi_j}^{1/2}\psi_j)
\label{kin}
\end{eqnarray}
where $Z_{\phi_i}$ ($Z_{\psi_j}$) and $Z_{\lambda_b}$ are the bosonic
(fermionic) wave function renormalizations and the coupling
renormalizations respectively, and $O_b(\phi_i,\psi_j)$ represent
interaction operators between the fields.  The $\widetilde\mu_i$
scales are appropriate combinations of  the independent scales $\mu$
and $\mu_i$, namely
\begin{equation}
\label{mutilde}
\log\widetilde\mu_i = \log\mu\,\theta(\mu-\mu_i)\ +\ \log\mu_i\, \theta(\mu_i-\mu) \
\end{equation}
Finally, the functions $f_{\phi_i}$, $f_{\psi_j}$ and $f_{\lambda_b}$
are dimensionless functions of the ratios
$\widetilde\mu_i/\widetilde\mu_j$ which are constant and finite and
can be expanded as: $f=1+{\cal O}(\hbar)$. They correspond to finite
wave-function and coupling renormalizations.  They should be chosen so
that the $\beta$ and $\gamma$-functions obtained from the bare
lagrangian  satisfy the decomposition given by Eqs.~(\ref{betamu}) and
(\ref{betamui}).

Let us notice that for $\mu<\mu_i$ (all $i$), we have
$\widetilde\mu_i=\mu_i$, while for $\mu>\mu_i$ (all $i$), we have
$\widetilde\mu_i=\mu$. In the latter region, we recover the ordinary
$\MSbar$-scheme. At intermediate values of $\mu$, the situation is
also intermediate: some of the $\widetilde\mu_i$ become equal to the
$\MSbar$-scale $\mu$.

It can be checked that the one-loop effective potential obtained  from
the lagrangian (\ref{lagdec}) has precisely the form of the  proposed
one-loop effective potential of Eq. (\ref{effpot}). In terms of the
$\widetilde\mu_i$ scales defined in Eq.~(\ref{mutilde}), it simply reads
\begin{equation}
\label{effpotn}
V_1=\frac{1}{64\pi^2} \sum_{i=1}^{N}(-)^{2s_i}
M_i^4\left[\log\frac{M_i^2}{\widetilde\mu_i^2}-C_i\right]
\end{equation}
In the next section we will apply this approach to our
simple extension of the Standard Model.

\nsect{Effective theory of the Standard Model extension}
\label{toy}

The presence of extra bosonic (fermionic) degrees of freedom tends to
stabilize (destabilize) the SM Higgs potential.  Consequently, when
the SM Higgs potential presents an instability at a certain scale,
$\Lambda_{SM}$, the most economic cure is the presence of just extra
bosonic fields.  Consequently, in this section we will apply the
results of section \ref{multiscale} to a very simple extension of the
Standard Model defined by the lagrangian
\begin{equation}
\label{lagtoy}
{\cal L}={\cal L}_{\rm
SM}+\frac{1}{2}\,\left(\partial_\mu\vec\Phi\right)^2 -
\frac{1}{2}\,\delta|H|^2\vec\Phi^2-\frac{1}{2}\,M^2\vec\Phi^2
-\frac{1}{4!}\lambda_s\vec\Phi^4
\end{equation}
where ${\cal L}_{\rm SM}$ is the SM lagrangian, $H$ the SM Higgs
doublet and $\vec\Phi$ $N$ real scalar fields transforming under the
vector representation of $O(N)$. $M$ is the invariant mass of
$\vec\Phi$, $\lambda_s$ its quartic coupling and $\delta$ provides the
mixing between the Higgs and $\vec\Phi$ fields. Since we are assuming
that $\vec\Phi$ is a SM singlet, $\delta$ is the only coupling that
involves the SM with the new physics and will play a relevant role in
our analysis.  Of course, this model may be unrealistic, but, for the
previous reasons, it basically represents the most efficent way to
cure a SM instability, thus giving reliable upper bounds on the scale
of the required new physics, i.e. the mass of the extra particles.

In the present case we have two relevant scales: the SM scale, which
can be conventionally chosen to be that corresponding to the
top-quark, $\mu_t$, and the scale of the new physics, $\mu_\Phi$.  In
the background of the Higgs field, $H^0=h_c/\sqrt{2}$, the one-loop
effective potential in the decoupling approach explained in the
previous section can be decomposed as in Eq.~(\ref{potencial}) with
\begin{equation}
\label{potarbol}
V_0=-\frac{1}{2}m^2 h_c^2+\frac{1}{4}\lambda h_c^4+\Lambda_c
\end{equation}
where $m^2$, $\lambda$ and $\Lambda_c$ are the SM mass term, quartic
coupling and vacuum energy, respectively, and $V_1$ 
as given by (\ref{effpot}) [or, equivalently, by (\ref{effpotn})]
\begin{eqnarray}
V_1&=&\frac{1}{64\pi^2}\left[ - 12
M_t^4(h_c)\left(\log\frac{M_t^2(h_c)}{\mu_t^2}
+\theta(\mu-\mu_t)\log\frac{\mu_t^2}{\mu^2}
-\frac{3}{2}
\right)\right.
\nonumber \\
&+& \left. N M_\Phi^4(h_c)\left(\log
\frac{M_\Phi^2(h_c)}{\mu_\Phi^2}
+\theta(\mu-\mu_\Phi)\log\frac{\mu_\Phi^2}{\mu^2}
-\frac{3}{2}\right)\right]\ .
\label{potuno}
\end{eqnarray}
The masses of the top-quark and the $\vec\Phi$ field are defined by
\begin{eqnarray}
\label{masas}
M_t^2(h_c)&=&\frac{\lambda_t\,h_c^2}{\sqrt{2}} \nonumber\\
M_\Phi^2(h_c)&=& M^2+\delta\,h_c^2\ ,
\end{eqnarray}
$\lambda_t$ being the SM top-quark Yukawa coupling. We are neglecting
in Eq.~(\ref{potuno}) the contribution to the one-loop effective
potential of all SM fields, except the top-quark field, as it is
usually done in SM studies.

To evaluate the potential given by Eqs.~(\ref{potarbol}, \ref{potuno})
we also need to know the ${\lambda_a}$ parameters (i.e. all the
masses, couplings and fields) at the corresponding values of the $\mu,
\mu_t, \mu_\Phi$ scales. Hence we need the  $\beta$ and
$\gamma$-function decomposition defined in (\ref{betamu}) and
(\ref{betamui}) for the relevant parameters of our model.  This is
given by
\begin{eqnarray}
_{\mu_t}\widetilde\beta_\lambda&=&\beta^{\rm SM}_\lambda \qquad \quad
_{\mu_\Phi}\widetilde\beta_\lambda =  \frac{2 N}{\kappa} \delta^2
\label{lambda}\\
_{\mu_t}\widetilde\beta_{m^2}&=&\beta^{\rm SM}_{m^2}
\qquad \quad  _{\mu_\Phi}\widetilde\beta_{m^2} = -\frac{2 N}{\kappa}
\delta M^2
\label{m2}\\
_{\mu_t}\widetilde\beta_{\Lambda}&=&\beta^{\rm
SM}_\Lambda  \qquad \quad _{\mu_\Phi}\widetilde\beta_{\Lambda}=
\frac{N}{2\,\kappa} M^4
\label{Lambda}\\
_{\mu_t}\widetilde\gamma_{h}&=&\gamma^{\rm SM}_h
\qquad \quad _{\mu_\Phi}\widetilde\gamma_{h} = 0
\label{gamma}
\end{eqnarray}
for the SM couplings, where $\kappa=16\pi^2$, and $\beta^{\rm SM}$,
$\gamma^{\rm SM}$ are the SM $\beta$ and $\gamma$-functions.  For
couplings corresponding to new physics the $\widetilde\beta$-functions
are:
\begin{eqnarray}
_{\mu_t}\widetilde\beta_{\delta}&=&2\gamma_h \delta+\frac{24}{\kappa}
\lambda \delta \nonumber\\
_{\mu_\Phi}\widetilde\beta_{\delta}&=&\frac{1}{\kappa}
\left(8\delta^2+\frac{1}{3}(N+2)\lambda_s \delta\right)
\label{delta}\\ && \nonumber\\
_{\mu_t}\widetilde\beta_{\lambda_s}&=& \frac{48}{\kappa} \delta^2
\nonumber\\ _{\mu_\Phi}\widetilde\beta_{\lambda_s}&=&
\frac{1}{3\,\kappa}(N+8)\lambda_s^2
\label{lambdas}\\ && \nonumber\\
_{\mu_t}\widetilde\beta_{M^2}&=&-\frac{8}{\kappa}\delta m^2
\nonumber\\
_{\mu_\Phi}\widetilde\beta_{M^2}&=&\frac{1}{3\,\kappa}\lambda_s (N+2)
M^2
\label{M2}
\end{eqnarray}

As noted in the previous section, and is apparent from
(\ref{potuno}),  for scales  $\mu>\mu_t,\mu_\Phi$, the effective
potential (\ref{potuno}) reduces to the ordinary $\MSbar$ effective
potential. However, for an improved evaluation of the potential it is
much more convenient to choose $\mu\simeq \mu_t\simeq M_t(h_c)$,
$\mu_\Phi\simeq M_\Phi(h_c)$, thus getting rid of dangerous logarithms. (In
the next section we will be more precise about the exact values of these
choices.) These values belong to the range $\mu_t\simlt \mu<\mu_\Phi$, where 
the one-loop effective potential can be written as:
\begin{eqnarray}
\label{potone}
V_1&=&V_1^{\rm SM}+ V_1^{\Phi} \nonumber\\ V_1^{\Phi}&=&
\frac{N}{64\,\pi^2} M_\Phi^4(h_c)\left(\log
\frac{M_\Phi^2(h_c)}{\mu_\Phi^2}-\frac{3}{2}\right)
\end{eqnarray}
Here $V_1^{\rm SM}$ is the usual SM one loop-effective potential in the
$\MSbar$-scheme, which depends explicitly on the RGE scale $\mu$,
while $V_1^{\Phi}$ corresponds to the contribution of the decoupled field $\vec\Phi$, which runs with $\mu_\Phi$.
Expressions (\ref{potarbol}, \ref{potone}) will be used in the next section to evaluate explicitly the effective potential.

To finish this section, it is interesting to write the explicit form
of the bare lagrangian in a multiscale renormalization scheme [see
Eqs.(\ref{lagdec}, \ref{kin})] which leads to the  effective potential
(\ref{potone}) and the set (\ref{lambda})--(\ref{M2}) of
$\widetilde\beta$-functions.  In the interesting range of scales,
$\mu_t\simlt \mu<\mu_\Phi$  [which, by Eq.~(\ref{mutilde}) implies
$\widetilde\mu_t=\mu$,  $\widetilde\mu_\Phi=\mu_\Phi$], this  is
explicitly given by
\begin{eqnarray}
\label{lagbare}
{\cal L}_{\rm Bare}&=&{\cal L}^{\rm SM}_{\rm Bare}\\
&+&\frac{1}{2}\mu_\Phi^{-\varepsilon/2} Z_\Phi  \left(\partial_\mu
\vec\Phi \right)^2-\frac{1}{2}Z_\Phi Z_{M^2} M^2
\vec\Phi^2-\frac{1}{4!} Z_\Phi^2 Z_{\lambda_s} \lambda_s \vec\Phi^4
-\frac{1}{2}Z_\Phi Z_H Z_\delta f_\delta \delta |H|^2 \vec\Phi^2
\nonumber
\end{eqnarray}
where ${\cal L}^{\rm SM}_{\rm Bare}$ is the SM bare lagrangian,  all
couplings and fields are renormalized, and all constant factors $Z_a$
have the form $Z_a=1+z_a/\varepsilon+\cdots$, where the $z_a$ factors
are those of the $\MSbar$ scheme.  The introduction of the finite
renormalization of the coupling $\delta$,
\begin{equation}
\label{rendelta}
f_\delta=\left(\frac{\mu}{\mu_\Phi}\right)^{4\,\delta/\kappa}
\end{equation}
is necessary in particular to satisfy the RGE given by
Eqs.~(\ref{delta}).

The term in $f_\delta$, when expanded to one-loop order,
$f=1+(4\,\delta/\kappa)\log\frac{\mu}{\mu_\Phi}+\cdots$ generates a
(finite counterterm)  contribution to the effective potential in the
presence of background fields $h_c$ and $\Phi$ as,
\begin{equation}
\label{finpot}
\Delta_c V(h_c,\Phi)=\frac{1}{16\,\pi^2}\delta^2\, h_c^2\, \vec\Phi^2
\log\frac{\mu}{\mu_\Phi}
\end{equation}
which grows logarithmically with the scales ratio $\mu/\mu_\Phi$.  In
principle, this is worrying, as it represents the kind of drawback of
the pure $\MSbar$-scheme in the presence of several scales that we
wanted to avoid with our approach. Besides, this seems to contradict the  form
of the effective potential (\ref{potone}), which is free  of such
dangerous logarithms. However, when turning the background field
$\Phi$ on and computing the one-loop diagram with external legs $|H|^2
\vec\Phi^2$ and internal propagators corresponding to a Higgs and a
$\vec\Phi$ field~\footnote{In fact this is the only non-trivial
one-loop diagram, in the sense that it contains internal lines
corresponding to the SM and to new physics. This diagram is
proportional to $\delta^2$ and plays a major role in our
construction.}, the contribution of the latter (after renormalization
in the $\MSbar$-scheme) precisely cancels that of the counterterm in
Eq.~(\ref{finpot}) and, altogether, the presence of the dangerous
$\log(\mu/\mu_\Phi)$ term, leading to the one-loop effective potential
given in Eq. (\ref{potone}).

\nsect{Numerical results}
\label{numerical}

In this section we will study the potential given by
Eqs.~(\ref{potarbol}, \ref{potone})  and analyze the conditions for
stability of the electroweak minimum at the vacuum expectation value
(VEV) $h_c=v=246$ GeV,  and the non-appearance of a (destabilizing)
deeper minimum  at larger values of the field. For given (fixed)
values of the Higgs VEV  $v$ and the Higgs mass squared $m_H^2$, the
minimum conditions of  the potential read\footnote{With the definition
of Eq.(\ref{minimo}) the physical (pole) squared Higgs mass $M_H^2$ 
is equal to $m_H^2$ plus some (small) radiative corrections,  which have been
taken into acount in the numerical computations.}
\begin{eqnarray}
\label{minimo}
\left. \frac{d\, V_{\rm eff}}{d\, h_c}\right|_{h_c=v}&=&0\nonumber\\
\left. \frac{d^2\, V_{\rm eff}}{d\, h^2_c}\right|_{h_c=v}&=&m_H^2
\end{eqnarray}
Using these conditions, the effective potential parameters, $m^2$  and
$\lambda$, can be traded by $v$ and $m_H^2$ as:
\begin{eqnarray}
\label{param}
m^2&=& m_{\rm SM}^2-\frac{N\,\delta^2}{\kappa}\,
v^2+\frac{N\,\delta}{\kappa} \,
M^2\left(\log\frac{M_\Phi^2}{\mu_\Phi^2}-1\right)\nonumber\\
\lambda&=& \lambda_{\rm SM}-\frac{N\,\delta^2}{\kappa}\log
\frac{M_\Phi^2}{\mu_\Phi^2}
\end{eqnarray}
where $m_{\rm SM}^2$, $\lambda_{\rm SM}$ represent the corresponding 
values as obtained in the pure SM
\begin{eqnarray}
\label{paramSM}
m^2_{\rm SM}&=& \frac{1}{2}\, m_H^2+\frac{3\, \lambda_t^4}{\kappa}\,
v^2 \nonumber\\ \lambda_{\rm SM}&=& \frac{m_H^2}{2\, v^2}+\frac{3\,
\lambda_t^4}{\kappa} \log\frac{M_t^2}{\mu^2}\ .
\end{eqnarray}

{}From the second equality in Eq.~(\ref{param}) we see that in order
to  preserve the validity of perturbation theory, a choice of the
scale,  $\mu_\Phi\simeq M_\Phi$, must be done, as expected.
Furthermore, from the first equality in Eq.~(\ref{param}),  we see
that for $M^2\gg m_H^2$ (which is the usual case) the third term  of
the right hand side will amount to a huge contribution, which must be
fine-tuned with the value of $m^2$, in order to keep the right scale
for $m_H$. This technical problem, which reflects a hierarchy
problem, is avoided by choosing $\mu_\Phi$ in such a way that the
annoying term in (\ref{param}) is cancelled, i.e.
\begin{equation}
\label{choice}
\log\frac{M_\Phi^2}{\mu_\Phi^2}=1 \ .
\end{equation}
Then the parameter fixing (\ref{param}) becomes,
\begin{eqnarray}
\label{paramf}
m^2&=& m_{\rm SM}^2-\frac{N\,\delta^2}{\kappa}\, v^2\nonumber\\
\lambda&=& \lambda_{\rm SM}-\frac{N\,\delta^2}{\kappa}\ .
\end{eqnarray}
Still perturbation theory can be spoiled, along with the SM vacuum,
for  values of $N$ and $\delta$ such that $N\delta^2 \simgt \kappa$ so
we will  restrict ourselves to the range of values such that
$N\delta^2 \ll \kappa$.

An immediate consequence of the choice (\ref{choice}) for $\mu_\Phi$
is that all the couplings of the theory, $\lambda_a$, acquire an
implicit $h_c$-dependence through their dependence on $\mu_\Phi$. In
fact this dependence can be obtained from the second equalities of
Eqs.~(\ref{lambda}--\ref{M2}) using
\begin{equation}
\label{betah}
\frac{d \lambda_a(h_c)}{d \log h_c}=
\left[1-\frac{M^2}{M_\Phi^2(h_c)}\right]\,
_{\mu_\Phi}\widetilde \beta_{\lambda_a}\theta(\mu_\Phi-\mu)
\ .
\end{equation}

We will consider now the effective potential given by
Eqs.~(\ref{potarbol})  and (\ref{potone}) and will improve it by the
RGE (\ref{lambda}) to  (\ref{M2}) and (\ref{betah}) with the choice
(\ref{choice}) of the $\mu_\Phi$ scale. The initial conditions for all
parameters will be taken at the boundary scales:
\begin{eqnarray}
\mu_0^2&=& M_t^2(v)\nonumber\\ \mu_{\Phi\,,0}^2&=& M_\phi^2(v)/e
\label{condin}
\end{eqnarray}
and those for $m^2$ and $\lambda$ will be fixed by (\ref{paramf}).
The system of partial differential equations (\ref{lambda}) to
(\ref{M2}) is solved by a step-wise procedure, which allows to
evaluate the effective potential for any value of $h_c$.
\begin{figure}[htb]
\centerline{
\psfig{file=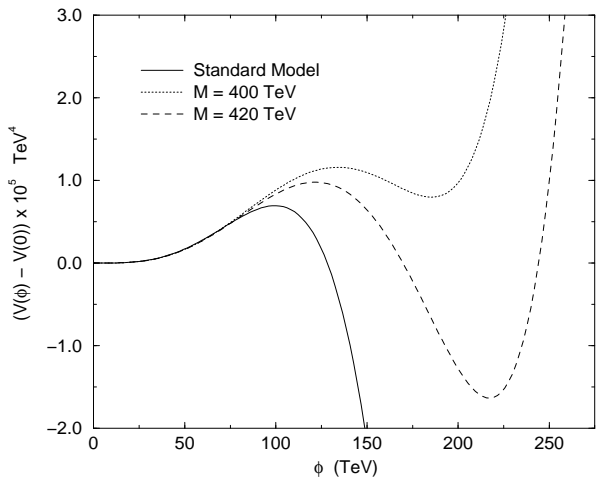,width=0.5\linewidth}}
\caption{Plot of the effective potential as function of the Higgs field
for $m_H=100$ GeV, $\delta=1$ and $N=10$.}
\label{fig1}
\end{figure}

In Fig.~\ref{fig1} we show a plot of the effective potential $V_{\rm
eff}$ for values of the Higgs and  top-quark masses, $m_H=100$ GeV and
$M_t(v)=175$ GeV, for the SM (solid line), which  shows an instability
for values of the field  $h_c\simeq 125$ TeV $\equiv\Lambda_{SM}$.
The dotted line corresponds to $V_{\rm eff}$ for the  SM extension
with $\delta=1$, $N=10$ and $M= 400$ TeV, which shows how the
instability is cured by the new physics. Smaller values of $M$ also
work.  Conversely, the SM results are recovered in the  limit when
$M\rightarrow\infty$ or $\delta\rightarrow 0$. This is illustrated by
the dashed line, which corresponds to $M=420$ TeV.
\begin{figure}[htb]
\centerline{
\psfig{file=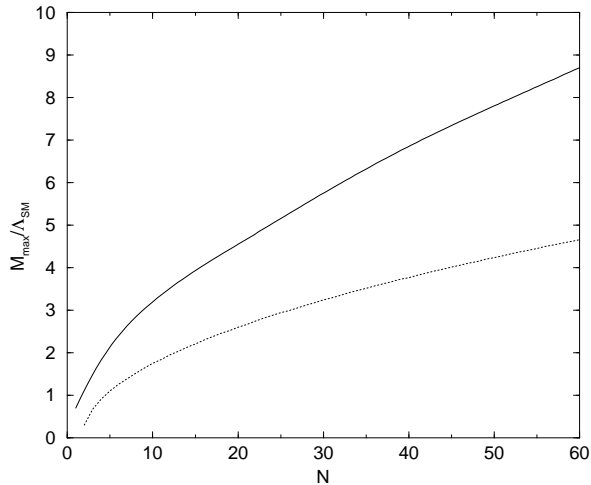,width=0.5\linewidth}}
\caption{Plot of the ratio  $M_{max}/\Lambda_{SM}$ for $\delta=1$ and 
$m_H=100$ GeV as function of the multiplicity $N$. The solid line and the 
dashed one shows the RGE improved and one-loop approximation 
respectively.}
\label{fig2}
\end{figure}

The previous example explicitely shows that the scale of new physics,
$M$, responsible for the cure of a SM instability, can be larger than
the scale $\Lambda_{SM}$ at which the SM instability develops. It
cannot be, however, arbitrarily larger. Fig.~\ref{fig2} shows (solid
line) the ratio $M_{max}/\Lambda_{SM}$ for $\delta=1$, as a function
of the number of extra degrees of freedom, $N$ (recall that $N\delta^2
\ll \kappa$ to preserve perturbativity). Clearly, $M$ could be as
large as $\simeq 10 \Lambda_{SM}$, which puts an upper bound on the
scale of new physics, $M$. For a typical value of the  multiplicity,
$N={\cal O}(10)$, we get the conservative bound  $M\simlt 4
\Lambda_{SM}$. This is e.g. the case of the MSSM, where the relevant
multiplicity is $N=12$, corresponding to the stops.  The dashed line
corresponds to the result of Ref.~\cite{sher2}, obtained in a cruder
approximation (one-loop instead of RGE improved). Our results confirm
the trend observed in that paper, but show that the ratio
$M_{max}/\Lambda_{SM}$ can be substantially larger than the one 
estimated there.
\begin{figure}[htb]
\centerline{
\psfig{file=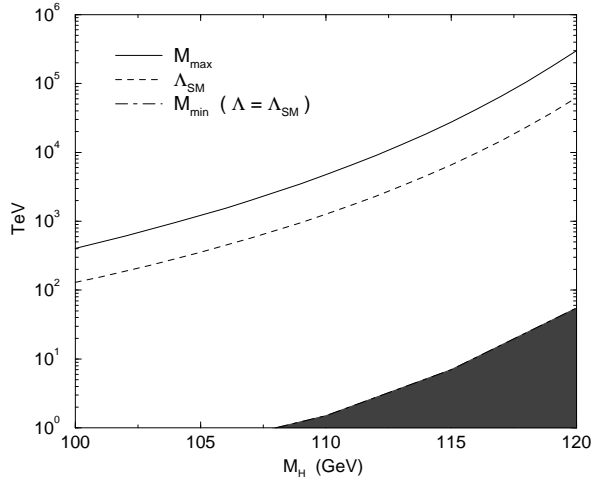,width=0.5\linewidth}}
\caption{Plot of $M_{max}$, $\Lambda_{SM}$ and $M_{min}$ as a 
function of the Higgs mass. The shadowed region shows the values not allowed 
for M.}
\label{fig3}
\end{figure}

Finally, Fig.~\ref{fig3} shows the smooth increasing of both,
$\Lambda_{SM}$ and $M_{max}$, with the Higgs mass, $M_H$. Also, there
appears a lower bound on $M$, coming from the requirement of
perturbativity up to the $\Lambda_{SM}$ scale.  This lower
bound may seem paradoxical. Actually, it is a feature of the
particular SM extension we have chosen: the lower $M$, the sooner the
new physics enters, which, due to the RGE (\ref{lambda}), may drive
more quickly the quartic Higgs coupling $\lambda$ into a
non-perturbative regime. Other SM extensions, in particular
supersymmetric extensions, do not present such lower limits on $M$. On
the other hand, the upper bound on $M$ is quite robust for any
conceivable SM extension, as has been explained at the beginning of
section 3.

The numerical results presented in Figs.~1--3 show the relation
between the scale $\Lambda_{SM}$, at which the SM develops a
instability, and the maximum value allowed for the scale of new physics
required to cure it. With the present experimental lower bounds on
$M_H$, $M_H\simgt 105$ GeV, it is clear that the possible new physics
could easily escape detection in the present and future accelerators.

\nsect{Conclusions}
\label{conclusions}

The possible detection of a relatively light Higgs ($M_H\simlt 140$
GeV) would imply an instability of the Higgs effective potential, thus
signaling the existence of new physics able to cure it. It is,
therefore, a relevant question what is the relation between the Higgs
mass (or, equivalently, the scale at which the instability develops,
$\Lambda_{SM}$) and the maximum allowed value of the scale of the new
physics, $M_{max}$.

In this paper we have examined this question in a rigorous way. This
requires, in the first place, a reliable approach to evaluate the
effective potential when several different mass scales are present. We
have followed the decoupling approach exposed in a previous paper
\cite{cdq1}, which can also be re-formulated as a multi-scale
renormalization approach, similar to those of Refs.~\cite{jones, ford}. Then, we
have considered a simple extension of the Standard Model, consisting
of $N$ extra scalar fields with an invariant mass $M$, coupled to the
Higgs field with a coupling $\delta$. This model, although
unrealistic, arguably represents the most efficent way to cure a SM
instability, thus giving reliable upper bounds on the scale of the
required new physics, i.e. the mass of the extra particles.

The numerical results,  presented in section 4, in particular in
Figs. 1--3, show the relation between $\Lambda_{SM}$ and
$M_{max}$. More precisely, for $\delta={\cal O}(1)$ and  $N={\cal
O}(10)$ (similar to the stop sector in the MSSM case), we obtain that
$M_{max}\simeq 4 \Lambda_{SM}$, which puts an upper bound on the scale
of new physics.  Unfortunately, the present lower bounds on the Higgs
mass, $M_H\simgt 105$ GeV, imply that $\Lambda_{SM}$ is at least
$10^2$ TeV. This fact, toghether with the previous upper bound on $M$,
implies that the new physics could easily (but not necessarily) escape
detection in the present and future accelerators.

\section*{ Acknowledgments}

We thank F. Feruglio, H. Haber and F. Zwirner for very useful discussions.


\end{document}